\documentclass[sigconf]{acmart}
\AtBeginDocument{%
  }

\newcommand{\tablestyle}[2]{\setlength{\tabcolsep}{#1}\renewcommand{\arraystretch}{#2}\centering\footnotesize}
\usepackage{tikz}
\usepackage{enumitem}
\usepackage{multirow}
\usepackage{pifont}
\usepackage{colortbl} 
\definecolor{COLOR_MEAN}{HTML}{f0f0f0}
\usepackage{amsmath}
\definecolor{LightCyan}{RGB}{232,241,255}
\definecolor{my_green}{RGB}{51,102,0}
\definecolor{my_red}{RGB}{204, 0, 0}
\renewcommand{\checkmark}{\textcolor{my_green}{\ding{51}}} 
\newcommand{\crossmark}{\textcolor{my_red}{\ding{55}}} 

\copyrightyear{2026}
\acmYear{2026}
\setcopyright{acmlicensed}
\acmConference[CIKM '26]{Proceedings of the 35th International ACM Conference on Knowledge and Information Management}{November 07--11, 2026}{Rome, ITALY}
\acmBooktitle{Proceedings of the 35th International ACM Conference on Knowledge and Information Management (CIKM '26), November 07--11, 2026, Rome, ITALY}

\begin{document}

\title{MuChator: Enabling Active Music Discovery via Conversational Music LLMs in Douyin Music}

\author{Jiahao Liang}
\authornote{Jiahao Liang, Linzhi Huang, and Xuannan Liu have equal contributions.}
\email{liangjiahao.24@bytedance.com}
\affiliation{%
  \institution{ByteDance}
  \city{Beijing}
  \country{China}
}

\author{Linzhi Huang}
\authornotemark[1]
\email{huanglinzhi.224488@bytedance.com}
\affiliation{%
  \institution{ByteDance}
  \city{Beijing}
  \country{China}
}

\author{Xuannan Liu}
\authornotemark[1]
\email{liuxuannan@bytedance.com}
\affiliation{%
  \institution{ByteDance}
  \city{Beijing}
  \country{China}
}

\author{Xukai Wang}
\email{wangxukai.v@bytedance.com}
\affiliation{%
  \institution{ByteDance}
  \city{Beijing}
  \country{China}
}

\author{Xuanpu Luo}
\email{luoxuanpu.01@bytedance.com}
\affiliation{%
  \institution{ByteDance}
  \city{Beijing}
  \country{China}
}

\author{Yongchun Zhu}
\authornote{Yongchun Zhu and Jingwu Chen are the corresponding authors.}
\email{zhuyongchun.zyc@bytedance.com}
\affiliation{%
  \institution{ByteDance}
  \city{Beijing}
  \country{China}
}

\author{Jingwu Chen}
\authornotemark[2]
\email{chenjingwu@bytedance.com}
\affiliation{%
  \institution{ByteDance}
  \city{Beijing}
  \country{China}
}

\author{Feng Zhang}
\email{feng.zhang@bytedance.com}
\affiliation{%
  \institution{ByteDance}
  \city{Beijing}
  \country{China}
}

\author{Xiao Yang}
\email{wuqi.shaw@bytedance.com}
\affiliation{%
  \institution{ByteDance}
  \city{Beijing}
  \country{China}
}

\renewcommand{\shortauthors}{Jiahao Liang et al.}

\begin{abstract}

Douyin Music, a large-scale platform with millions of daily users, adopts an immersive, feed-based discovery paradigm, where users passively explore music through continuous recommendations. 
While effective for \textbf{passive music discovery}, this paradigm restricts users to recommendation results and provides limited support for explicitly specifying listening intents. 
Unlike conventional search, where users express well-defined intents through explicit queries such as specific songs or artists, real-world \textbf{active music discovery} is often situational and colloquial, involving vague or underspecified requests, e.g., seeking music suitable for falling asleep.
While LLMs enable natural language interaction, their direct use in music discovery remains limited by insufficient \textit{music-domain knowledge}, the absence of \textit{music-query collaborative reasoning}, and a shallow understanding of \textit{personalized preferences}.
To address these challenges, we introduce \textbf{MuChator}, an interactive MusicLLM-based framework that enables users to actively express situational music intents in natural, colloquial language. 
MuChator incorporates three key components:
(1) \textbf{Music Knowledge Pre-training}, a three-stage pre-training scheme that incrementally incorporates objective music knowledge, subjective music knowledge, and personalized music preferences into LLMs;
(2) \textbf{Context-aware Instruction Tuning}, which constructs high-quality user–query–music triplets through an automated synthesis pipeline to align LLMs with active and situational user intents; and
(3) \textbf{Preference Alignment with Hybrid RM}, which jointly models intent relevance, personalized preferences, and basic constraints, and is optimized using GRPO-based reinforcement learning.
Extensive evaluations on industrial music recommendation datasets demonstrate that MuChator outperforms leading proprietary models, such as Gemini-3-Pro. The model has been deployed on Douyin Music App within ByteDance, with 46.49\% improvement of user active days in online A/B test.

\end{abstract}




\maketitle

\section{Introduction}

\begin{figure}[!t]
  \centering
    \includegraphics[width=0.96\linewidth]{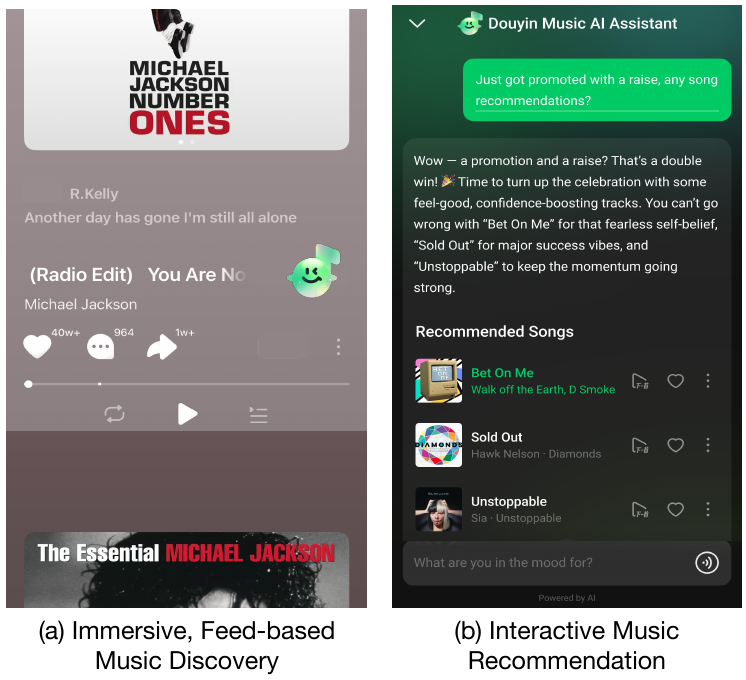}
    \vspace{-1.0em}
    \caption{ (a) Immersive, feed-based music discovery, where users passively explore music through continuous recommendations. (b) Interactive music recommendation that enables users to actively express situational music intents in natural, colloquial language.
    }
    \label{fig:figure1}
\end{figure}

Douyin Music, one of the largest music platforms in China, serves millions of daily active users. As shown in Figure~\ref{fig:figure1} (a), this platform emphasizes an immersive, feed-based music discovery paradigm, where music is continuously delivered in a stream-like manner similar to Douyin and TikTok. 
Such paradigm allows users to seamlessly switch the app to the background, thereby encouraging prolonged and passive consumption~\cite{schedl2018current}. Empirical results show that over 90\% of user listening time occurs in background playback. Most existing music recommendation methods~\cite{sanchez2016collaborative,shakirova2017collaborative,zhu2024interest,jing2025emotion} focus on improving the performance of passive music recommendation and have made significant progress.

Our findings suggest that passive music discovery alone cannot satisfy users’ immediate music needs. At app launch, users often engage in proactive music seeking instead of directly entering background listening. Once satisfactory music is found, users transition to immersive listening; otherwise, the listening session often terminates early. For active music discovery, users mainly rely on two interface components: radio channels and music search. However, both present notable limitations. (1) Radio channels are organized by predefined genre tags (e.g., metal music), requiring specialized music knowledge that many users do not possess. (2) Music search relies on semantic retrieval and primarily supports well-defined intents, such as explicit queries for specific songs or artists.

However, due to the lack of explicit music knowledge and clearly defined intents, active music discovery is often situational and colloquial, with users issuing vague or underspecified requests, e.g., ``songs suitable for a walk by the seaside''.
Therefore, a novel interactive music recommendation paradigm that enables convenient and active music discovery is essential. A promising paradigm for interactive music recommendation leverages natural-language interfaces to enable users to actively express open-ended listening intents. As shown in Figure~\ref{fig:figure1} (b), a user can issue a conversational request (e.g., ``Just got promoted with a raise, any song recommendations''), and the assistant responds with a curated playlist (e.g., ``Bet On Me'', ``Sold Out'', and ``Unstoppable'') aligned with the expressed intent. By enabling real-time conversational interaction, this paradigm mitigates the rigidity of traditional interface components and provides a more expressive pathway for music discovery.

Large Language Models (LLMs)~\cite{yang2025qwen3, liu2024deepseek, chatgpt2025, comanici2025gemini} provide a natural foundation for building conversational music recommendation assistants.
However, despite their proficiency in general-purpose instruction following and semantic reasoning, directly applying off-the-shelf LLMs to music recommendation presents three main challenges. \textbf{(1) Insufficient Music-domain Knowledge}. LLMs are not inherently tailored for music recommendation and lack specialized knowledge grounded in large-scale music catalogs, including songs, artists, and platform-specific item identifiers. \textbf{(2) Absence of Music-query Collaborative Reasoning}. Beyond semantic relevance derived from music-query alignment, relevance additionally requires reasoning over user context and music-query collaborative relationships.
Such capability is typically absent in LLMs trained on general-purpose corpora, especially when interpreting abstract queries with high-level semantics.
\textbf{(3) Shallow Understanding of Personalized Preferences}. 
Generic LLMs are insensitive to user-specific preferences derived from historical listening behaviors. As a result, they struggle to capture implicit user interests.


To address these challenges, we propose \textbf{MuChator}, an end-to-end LLM-based intelligent assistant for conversational music recommendation at Douyin Music APP of ByteDance.
MuChator is designed to integrate a conversational interaction with a personalized engine, enabling user-centric music discovery.
At its core, we first propose a three-stage Music Knowledge Pre-training which employs a curriculum learning strategy to augment LLMs with objective music knowledge, subjective music knowledge, and  personalized music preference.
To facilitate preference learning, we propose a next-behavior prediction paradigm that unifies item recommendation and user feedback modeling under an autoregressive objective.
Moreover, we introduce a Context-aware Instruction Tuning to align LLMs with active and situational user intents by synthesizing high-quality user–query–music triplets. These triplets are constructed through an automated pipeline that integrates semantic clustering, candidate retrieval, and personalization-guided filtering.
Finally, we introduce Preference Alignment with hybrid reward models (RM), optimized via GRPO-based reinforcement learning. These reward models are designed to jointly model intent relevance, personalized preferences, and basic constraints.

Our main contributions include: \textbf{(1)} We propose MuChator, an interactive MusicLLM-based assistant deployed in Douyin Music App, offering a new interaction alternative for music recommendation. \textbf{(2)} We introduce three-stage Music Knowledge Pre-training to internalize objective music knowledge, subjective music knowledge, and personalized music preference. To support preference learning, we propose next-behavior prediction that unifies music recommendation and feedback modeling. \textbf{(3)} We introduce a Preference Alignment Post-training that constructs User–Query-to–Music triplets for instruction tuning, and optimizes hybrid rewards via GRPO-based reinforcement learning. \textbf{(4)} Extensive experiments validate MuChator’s effectiveness and demonstrate its advantages against state-of-the-art LLMs in online and offline settings.

\begin{figure*}[!ht]
  \centering
    \includegraphics[width=0.98\linewidth]{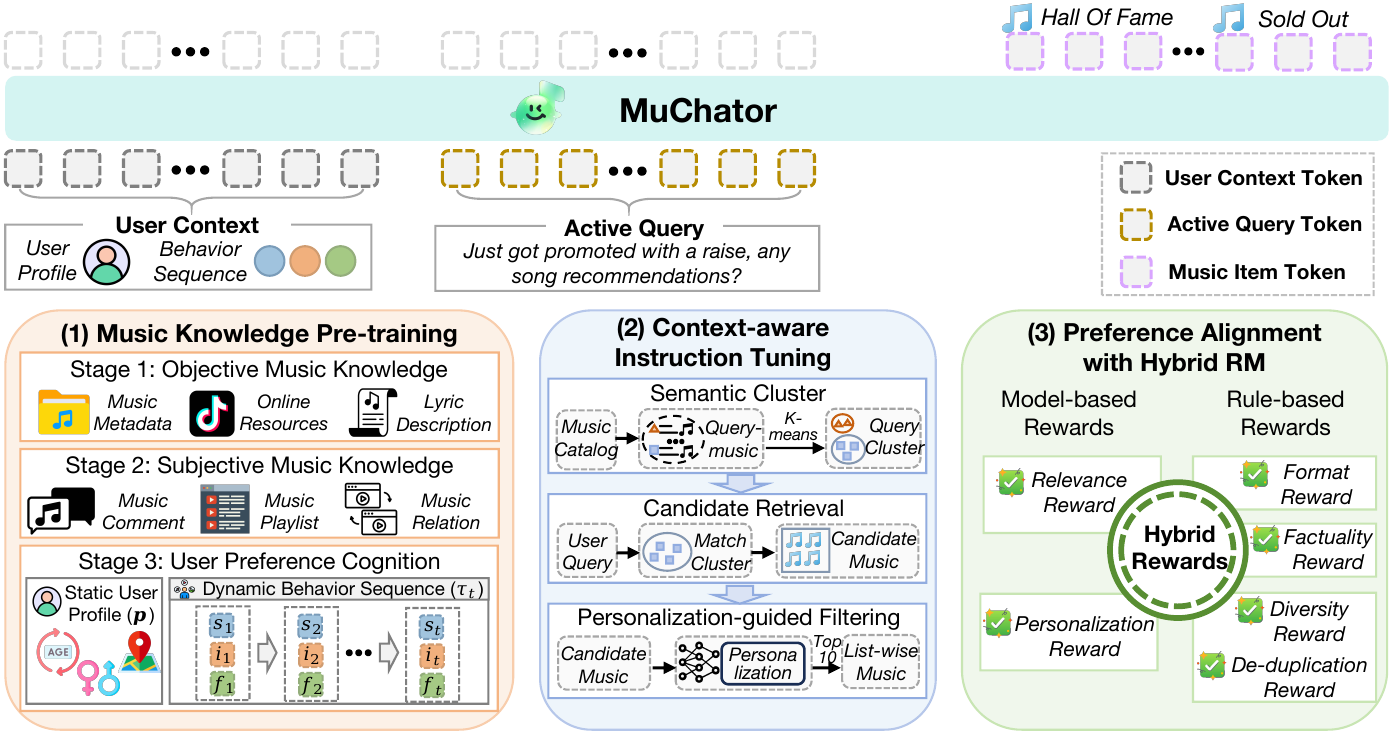}
    \vspace{-0.6em}
    \caption{ The framework of the MuChator. We first propose a three-stage Music Knowledge Pre-training to progressively internalize objective music knowledge, subjective music knowledge, and personalized music preference. We then introduce Context-aware Instruction Tuning to adapt the model for reasoning from user context and user queries to music items. Finally, we devise Personalization-driven GRPO by integrating hybrid rewards for intent and preference alignment.
    }
    \label{fig:framework}
\end{figure*}

\section{Related Work}
\noindent  \textbf{Music Recommendation.}
Traditional music recommender systems aim to suggest tracks that align with users’ preferences, typically relying on collaborative filtering (CF) methods, content-based methods, and context-aware methods.
CF methods~\cite{sanchez2016collaborative,shakirova2017collaborative} recommend music by measuring similarities derived from user–item interactions, either among users based on listening histories or among tracks based on play records.
Content-based methods~\cite{van2013deep, wang2014improving,deldjoo2024content,jin2023order} mainly focus on analyzing item attributes, including acoustic features (e.g., melody and rhythm) and textual metadata (e.g., tags and lyrics).
To integrate various types of information, context-aware methods~\cite{wang2021modeling,wang2023multi} have been proposed to incorporate users’ contextual cues related to static profiles, situational state (e.g., time, location, weather, activity and emotion)~\cite{jing2025emotion,shen2020peia, liu2023emotion, he2025eeg}, listening behavior~\cite{wang2020came} and feedback~\cite{oramas2016sound}. 

To address the demand of user-initiated queries, LLMs have been adopted for music recommendation~\cite{epure2025music}.
Some works leverage LLMs in recommendation either as feature extractors to provide semantic representations~\cite{DBLP:conf/kdd/WangOZXRPLL25} or as data generators to increase diversity~\cite{melchiorre2025just}. While effective, these methods largely overlook users’ interactive needs through natural language conversations. Recent works either to exploit the in-context learning abilities of LLMs~\cite{doh2025talkplay-tool} or adapt them to the music domain via continued pre-training on large-scale music corpora~\cite{tian2025mucpt,doh2025talkplay}. 
However, these works lack explicitly modeling preference alignment in intent relevance, personalized preferences throughout the training stages, which limits their ability to tackle complex user intents.
In contrast, we propose MuChator, an interactive MusicLLM-based framework that systematically integrates music knowledge pre-training, context-aware instruction tuning, and preference alignment with hybrid RM.

\noindent \textbf{Conversational Recommendation.}
Conversational recommendation systems aim to engage users in natural language dialogues to deliver personalized recommendations~\cite{christakopoulou2016towards}. 
Early methods use 
two separate modules for conversation and recommendation~\cite{chen2019towards,li2021seamlessly,zhou2020improving, an2025beyond,guan2025effective}, resulting the issue of semantic inconsistency~\cite{wang2022towards}.
Recently, the extensive knowledge and dialogue capabilities of LLMs have enabled to unify these modules into a cohesive textual space.
One line of work explores LLMs’ recommendation capability in the zero-shot setting via prompt engineering~\cite{he2023large,zhu2025collaborative}. 
However, their performance is hindered by the inherent domain gap between the knowledge acquired from general corpora and the specific user-item interaction patterns in recommendation scenarios~\cite{dai2023uncovering}. To address this issue, more recent works~\cite{dao2024broadening, zhang2025recommendation, yang2025step, xi2025bursting, shan2025automatic, lu2025roma} fine-tuning LLMs using recommendation data, enabling them to better comprehend recommendation-specific contexts.

\section{Methodology}
\subsection{Overview}
As shown in Figure~\ref{fig:framework}, MuChator consists of three main stages. 
First, Music Knowledge Pre-training utilizes curriculum learning to internalize domain knowledge spanning objective music knowledge, subjective music knowledge and personalized music preference.
Second, in Context-Aware Instruction Tuning, we construct User–Query–to–Item (UQ2I) instruction data to align the model’s list-wise outputs with active queries and user contexts.
Finally, in Hybrid Rewards for Reinforcement Learning (RL), we design a hybrid reward that integrates model-based feedback and rule-based constraint, and optimize the model via a GRPO-based algorithm.

\subsection{Music Knowledge Pre-training}
Music recommendation differs markedly from general LLM pre-training in its underlying corpus distribution.
To bridge this gap, we introduce Music Knowledge Pre-training, which progressively equips the model with broad knowledge of musical concepts, collaborative reasoning over query-music relevance, and precise alignment with personalized tastes.
To facilitate preference learning, we propose a novel paradigm of next-behavior prediction that unifies music item recommendation and user feedback modeling. 


\begin{table}[!t]
  \centering
  \caption{Statistics of datasets used in the three-stage Music Knowledge Pre-training, summarized in terms of dataset type, dataset description, and token cost.}
       \vspace{-1.0em}
  \label{data_distribution}
    \resizebox{1.0\linewidth}{!}{
    \tablestyle{4.0pt}{1.0}
   \begin{tabular}{c|c|c}
\toprule
\textbf{Dataset Type}       & \textbf{Dataset Description}                  & \textbf{Token} \\ \midrule

\rowcolor{COLOR_MEAN}
\multicolumn{3}{l}{\textit{Stage 1: Objective Music Knowledge}}                   \\
Music Metadata & title, singer,  lyric, genre & 1B \\
Online Resource                & encyclopedia, articles and background                            & 1.2B     \\
Lyric Description               & Interpret lyrics into descriptions                         & 0.5B    \\ \midrule
\rowcolor{COLOR_MEAN}
\multicolumn{3}{l}{\textit{Stage 2: Subjective Music Knowledge}}                  \\
Music Comment     &     user comments and professional reviews       &   2B   \\
Music Playlist    &          user playlist     &    0.8B  \\ 
Music Relation    &      item-cf data and music knowledge graphs  &  1.8B    \\              \midrule
\rowcolor{COLOR_MEAN}
\multicolumn{3}{l}{\textit{Stage 3: Personalized Music Preference}}                  \\
User Context      &  User Profile, User Behavior Sequence     &   16B    \\
               \bottomrule
\end{tabular}
}
\end{table}

\subsubsection{Three-stage Music Knowledge Pre-training}
Music Knowledge Pre-training adopts a curriculum learning strategy~\cite{liu2025incomplete} that progressively injects LLMs with three types of music knowledge in increasing order of difficulty: (1) objective music knowledge to ground fundamental musical concepts, (2) subjective music understanding to capture dependencies for query-music reasoning, and (3) personalized music preference to learn user behavior patterns.

\begin{figure}[!t]
  \centering
    \includegraphics[width=0.98\linewidth]{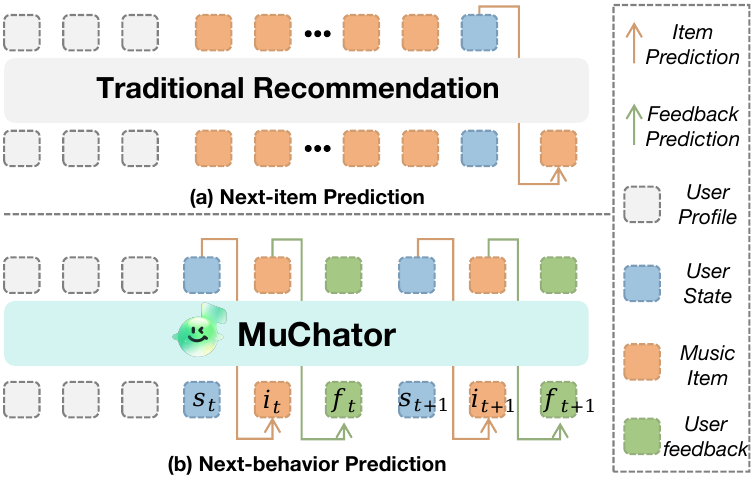}
    \vspace{-1.0em}
    \caption{ Comparison of sequence modeling paradigms in recommendation. (a) Next-item prediction calculates the loss solely on the final item, resulting in low training efficiency. (b) Next-behavior prediction interleaves item and feedback tokens for joint prediction, providing denser supervision.
    }
    \vspace{-1.0em}
    \label{fig:next-behavior-prediction}
\end{figure}

\begin{itemize}[noitemsep,left=2pt, itemsep=0pt, topsep=0pt]

\item \textbf{Stage 1: Objective Music Knowledge.}
We first construct a large-scale general music corpus with high quality and broad coverage. As shown in Table~\ref{data_distribution}, the corpus include three data types to characterize music concepts in a complementary manner:
\textbf{(1) Music Metadata}. We collect structured music metadata (i.e., song titles, artists, lyrics, and genres) from Douyin Music. These metadata are inserted into different templates to form training instances.
\textbf{(2) Online Resources}. To provide rich semantics beyond raw metadata, we collect music-related textual documents from Douyin platform, such as encyclopedia, song-related articles, and cultural background etc.
\textbf{(3) Lyric Description}. Since Lyrics often convey abstract semantics, we leverage LLMs to interpret lyrical content into detailed descriptions. 

\item \textbf{Stage 2: Subjective Music Knowledge.}
To support collaborative reasoning over query-music relevance, we incorporate three types of data to equip the model with subjective music knowledge, as summarized in Table~\ref{data_distribution}:
\textbf{(1) Music Comments}. We collect user comments and professional reviews from the Douyin Music platform to build connections between subjective musical discernment and objective music attributes.
\textbf{(2) Music Playlists}. We incorporate large-scale playlist data to capture collective listening behaviors and item co-occurrence patterns.
\textbf{(3) Music Relations}. We further incorporate structured relational signals, including CF–based item associations and music knowledge graphs linking songs and comments.

\item \textbf{Stage 3: Personalized Music Preference.}
To internalize user preference patterns, we further extend the pre-training phase by integrating rich user context data. This data is collected from Douyin Music and includes user profiles (e.g., age, gender, and occupation) as well as behavior sequences of real-time states, items, and  feedback.
To model user context sequences, as shown in Figure~\ref{fig:next-behavior-prediction} (a), traditional methods~\cite{deng2025onerec, sun2019bert4rec, hidasi2015session} adopt next-item prediction, which predicts the next item based on the user profile and historical music playlist. While effective, this paradigm calculates the loss solely at the final token of each sequence, resulting in low training efficiency.
\end{itemize}

\subsubsection{Next-behavior Prediction.} 
To address this problem, as shown in Figure~\ref{fig:next-behavior-prediction} (b), we propose a novel paradigm that formulates user behaviors as time-series trajectories and interleaves item and feedback prediction within an unified autoregressive sequence.
\begin{itemize}[noitemsep,left=2pt, itemsep=0pt, topsep=0pt]

\item \textbf{Time-series User Context.} The user context $\mathcal{C}_t$ at time $t$ is formulated as a combination of a static user profile $\mathbf{p}$ and a dynamic user behavior sequence $\tau_t$.
The user behavior $\tau_t$ is an interleaved sequence of situational states $s_t$, listening tracks $i_t$, and feedback signals $f_t$, which can be written as:
\begin{equation}
    \tau_t = (s_1, i_1, f_1,...,s_t, i_t, f_t).
\end{equation}

\item \textbf{Unified Autoregressive.} To precisely capture user preference information, we identify two tasks: item prediction (what to recommend) and feedback prediction (how users respond).
\textbf{(1) Item Prediction}. Given a previous user context $\mathcal{C}_t$ and a current user state $s_{t+1}$, the model learns a probability distribution over the candidate item space to predict the next interacted item $i_{t+1}$:
\begin{equation}
\max_{\theta}\ \sum_{t}\Big[
\log p_{\theta}\left(i_{t+1}\mid \mathcal{C}_t, s_{t+1}\right)\Big].
\end{equation}
\textbf{(2) Feedback Prediction}. While item prediction identifies which track a user is likely to listen, it fails to capture the user's feedback to the music content. We thus introduce a feedback prediction task, which conditions on both the user context and the predicted next item $i_{t+1}$ to estimate the feedback label $f_{t+1}$ (e.g., like, skip, dislike):
\begin{equation}
    \max_{\theta}\ \sum_{t}\Big[
\log p_{\theta}\left(f_{t+1}\mid \mathcal{C}_t, s_{t+1}, i_{t+1}\right)\Big].
\end{equation}
The item and feedback prediction tasks are interleaved within a single sequence and optimized under a unified autoregressive objective. This joint formulation enables the model to learn both what to recommend and how users respond in one framework. 

\end{itemize}

\subsection{Context-aware Instruction Tuning}
To support effective alignment between user queries and list-wise item recommendations, we introduce supervised fine-tuning tailored for interactive music recommendation. However, high-quality query–item pairs, particularly those grounded in rich user context, are scarce in practice. To address this problem, we introduce a automated pipeline for synthesizing User–Query–to–Item (UQ2I) instruction data, where each sample takes the user context and query as inputs, and a list-wise music items as output.



\subsubsection{Construction of UQ2I Data.}
As shown in Figure~\ref{fig:framework}, the automated pipeline for generating UQ2I instruction data consists of three steps:
\textbf{Step 1: Semantic Clustering.}
We first construct a query–item pool by integrating two complementary data sources: (1) leveraging LLMs to generate diverse natural language queries conditioned on individual songs; and (2) incorporating real-world online queries to enhance data authenticity and coverage. We then perform semantic clustering over the collected queries, yielding approximately 40K query clusters, each representing a distinct class of musical intent.
\textbf{Step 2: Candidate Retrieval.}
Given a specific user, we identify a set of songs that the user has previously favorited based on historical interactions. For each saved song, we randomly sample a query from its associated query list and determine its corresponding query cluster. Then, all songs indexed within this cluster are retrieved to constitute the candidate songs.
\textbf{Step 3: Personalization-guided Filtering.}
For candidate songs, we apply a fine-tuned personalization reward model (detailed in Section~\ref{hybrid_reward}) to assign personalized scores to each song. Finally, we select the top-10 highest-scoring songs to form the list-wise recommendation output.

\subsubsection{Training Objective.} Given the user context $\mathcal{C}$ and query $q$ from the collected UQ2I instruction data $\mathcal{D}_{\text{UQ2I}}$, the model is trained to generate a ranked list of music items $\mathbf{y} = (y_1, \dots, y_T)$ with the objective
$\mathcal{L}_{\text{SFT}} = - \mathbb{E}_{(\mathcal{C}, q, \mathbf{y}) \sim \mathcal{D}_{\text{UQ2I}}} \sum_{t=1}^{T} \log p_{\theta}(y_t \mid y_{<t}, \mathcal{C}, q)$.

\subsection{Preference Alignment with Hybrid RM}
After supervised fine-tuning, we further introduce personalization-driven reinforcement learning to achieve fine-grained alignment in user preferences and query–item relevance. 
Specifically, the reinforcement learning stage is implemented via Group-wise Policy Optimization (GRPO) with carefully designed hybrid rewards.
To mitigate overfitting, we adopt a strict data isolation strategy by constructing the reinforcement learning dataset using only users and queries that are unseen during supervised fine-tuning.

\subsubsection{Construction of Hybrid Rewards.} 
\label{hybrid_reward}
We design a hybrid reward strategy that integrates two complementary components: model-based rewards and rule-based rewards, as detailed below.

\begin{itemize}[noitemsep,left=2pt, itemsep=0pt, topsep=0pt]

\item \textbf{Model-based Rewards} leverage off-the-shelf models to quantify query-item relevance and personalized preferences.
\textit{(1) Relevance Reward} ($R_{\text{rel}}$).
We employ a fine-tuned LLM as a judge to evaluate semantic relevance between user queries and generated music items.
The judge model is trained via supervised fine-tuning on a curated instruction dataset derived from high-quality song summaries, enabling reliable relevance discrimination.
During the reinforcement learning stage, the resulting relevance reward is binarized into a hard gating signal indicating query–item relevance or irrelevance.
This design enforces strict alignment with user intent, encouraging the model to prioritize query understanding.
\textit{(2) Personalization Reward} ($R_{\text{pers}}$).
We use the online multi-objective ranking model as the personalized reward evaluator to fit real user feedback. With large-scale training on user behavior data from the Douyin Music App, this model can effectively quantify preference-level alignment between candidate music items and user context.
Consequently, it assigns higher rewards to recommendations that exhibit continuity with users’ established interests, yielding a personalization reward.

\item \textbf{Rule-based Rewards} are designed to impose practical constraints in real-world recommendation scenarios.
\textit{(1) Format Reward} ($R_{\text{format}}$) is employed to ensure valid parsing of music items. 
\textit{(2) Factuality Reward} ($R_{\text{fact}}$) is used to verify that the generated items exist in the platform's catalog and prevent hallucinations. 
\textit{(3) Diversity Reward} ($R_{\text{div}}$) aims to penalize repeated artists or duplicated music title suffixes within the recommendation list.
\textit{(4) De-duplication Reward} ($R_{\text{dedup}}$) is designed to discourage recommending items that appear in the user’s recent positive interaction history.
\end{itemize}
Based on the aforementioned six rewards, we construct a hybrid reward function $R_{\text{hyb}}$, calculated as
$R_{\text{hyb}} = R_{\text{rel}} \cdot \lambda_{\text{pers}} R_{\text{pers}} + \sum \lambda_{\text{rule}} R_{\text{rule}}$.
Here, $\mathcal{R}_{\text{rule}} = \{R_{\text{format}}, R_{\text{fact}}, R_{\text{div}}, R_{\text{dedup}}\}$ denotes the set of rule-based rewards, $R_{\text{rel}} \in \{0,1\}$ is a binary gating signal, and $\lambda_{\text{pers}}$ and $\lambda_{\text{rule}}$ are weighting coefficients.

\subsubsection{Training Objective.} 
Given the user context $\mathcal{C}$, query $q$, and a sampled group of candidate recommendation lists $\{\mathbf{y}_{i}\}_{i=1}^{N}$ from the current policy $p_{\theta}$, the objective maximizes the expected hybrid reward while constraining the policy update:
\begin{equation}
\begin{aligned}
\mathcal{L}_{\text{GRPO}}
=
- \mathbb{E}_{(\mathcal{C}, q)\sim\mathcal{D}}
\Bigg[
&\frac{1}{G}\sum_{i=1}^{G}
\min\Big(
r_i A_i,\,
\mathrm{clip}\big(r_i, 1-\epsilon, 1+\epsilon\big) A_i
\Big) \\
&- \beta\,
D_{\mathrm{KL}}\!\Big(
p_\theta
\,\big\|\,
p_{\text{ref}}
\Big)
\Bigg].
\end{aligned}
\end{equation}
where the importance ratio $r_i$ is calculated by
$r_i = \frac{p_\theta(\mathbf{y}_i \mid \mathcal{C}, q)}{p_{\text{ref}}(\mathbf{y}_i \mid \mathcal{C}, q)}$
, and the advance value $A_i$ is calculated by $A_i = R_i - \frac{1}{G} \sum_{j=1}^{G} R_j$.

\section{Experiments}

\subsection{Experimental Setup}

\subsubsection{Evaluation Datasets and Online Environment.}
\textbf{Datasets.} 
We evaluate MuChator with baseline models on a large-scale industrial recommendation dataset, DouyinMusic-MuChator. 
DouyinMusic-MuChator is constructed following the same design principles as the DouyinMusic-20B~\cite{zhu2024interest,zhu2025long}, containing over 20,000 samples collected from the impression logs of Douyin Music App.
Each sample in the industrial dataset is collected across the time span of 4 weeks in December 2025.
\textbf{Online Environment.} All online experiments are conducted on Douyin Music’s primary online system. The online A/B testing environment supports large-scale experiments, with daily metrics including daily users over 0.1 billion.

\subsubsection{Evaluated Baselines.}
We compare MuChator against three groups of baselines: 
(1) zero-shot baselines using two frontier proprietary LLMs, OpenAI's GPT-5.2 and Google's Gemini-3-Pro; 
(2) few-shot baselines using the same models, with five carefully curated in-context examples; and 
(3) tuning-based baselines built on Qwen-3. 
Specifically, we train a supervised fine-tuning (SFT) baseline using the same supervision data as MuChator, and also include retrieval-augmented generation (RAG) baselines tuned on Qwen-3.

\subsubsection{Evaluated Metrics.}
\textbf{Offline Evaluation.} 
We evaluate results across four metrics: 
\textbf{(1) Personalization} is evaluated using the online multi-objective ranking model to obtain preference scores;
\textbf{(2) Relevance} is evaluated with the fine-tuned relevance reward model to obtain query-to-item relevance scores;
\textbf{(3) Diversity} measures the proportion of distinct music recommended by the model in the Douyin music catalog.
\textbf{(4) Factuality} is measured by the proportion of recommended music items matched with the Douyin Music catalog.
\textbf{Online Evaluation.} 
Online evaluation measures user involvement using three metrics: (1) Active Day measures the average number of days a user plays music; (2) Duration measures average duration of music playback per user. (3) Click-Through Rate (CTR) measures the ratio of user clicks to recommended messages.


 \subsubsection{Implementation Details.} 
We utilize Qwen3-8B~\cite{yang2025qwen3} as the backbone model, initialized from publicly available pre-trained weights.
For training hyperparameters, we use a packing token size of 7,168 and a learning rate of $2 \times 10^{-5}$ for one epoch in pre-training, a batch size of 256 and a learning rate of $2 \times 10^{-5}$ for two epochs in SFT, and a batch size of 1,024, a rollout sample size of 8, and a learning rate of $1 \times 10^{-5}$ for one epoch in RL. In deployment, MuChator achieves a 0.09s time-to-first-token and 1.87s end-to-end latency for 140-token average responses.
All experiments are performed on a cluster of 512 flagship GPUs.

\subsection{Main Results}

\begin{table}[!tt]
  \centering
  \caption{Online A/B testing results. Each row indicates the relative improvement over the internal music search system.}
    \vspace{-1.2em}
  \label{online_result}
    \resizebox{1.0\linewidth}{!}{
    \tablestyle{4.0pt}{1.0}
\begin{tabular}{l|ccc}
\toprule
\rowcolor{COLOR_MEAN}
\textbf{Method}  & \multicolumn{1}{c}{\textbf{Active Day$\uparrow$}} & \multicolumn{1}{c}{\textbf{Duration$\uparrow$}} & \multicolumn{1}{c}{\textbf{CTR$\uparrow$}}  \\ \midrule

Qwen3-8B$_{\text{SFT}}$     &  -5.81\%   &  -9.21\%    &   -2.42\%   \\
\rowcolor{LightCyan}
\textbf{MuChator}      &  +46.49\%   &  +77.36\%    &   +11.26\%   

\\ \bottomrule
\end{tabular}
 \vspace{-1.2em}
}
\end{table}

\subsubsection{Online A/B Testing.} 
To assess the platform revenue of MuChator, we conducted an online A/B test on Douyin Music App for approximately one month. 
We evaluated the online performance of two models, MuChator and Qwen3-8B fine-tuned with SFT, using Active Day, Duration, and Click-Through Rate (CTR).
The online A/B results are presented in Table~\ref{online_result}. 
The online baseline is an internal music search system, i.e., \textit{a non-LLM recommender that combines retrieval and ranking stages}, leveraging Douyin Music data for continual improvement. 
MuChator achieves consistent improvements in all three key metrics for all users: Active Day (+46.49\%), Duration (+77.36\%), and CTR (+11.26\%) while SFT on Qwen3-8B alone does not yield online gains.
These results demonstrate that MuChator, which integrates domain-specific pre-training and post-training, can proactively satisfy users' immediate music intents, thereby substantially improving user experience.

\begin{table}[!t]
  \centering
  \caption{Offline experiment results of comparison MuChator with three groups of baselines. }
    \vspace{-1.2em}
  \label{main_result}
    \resizebox{1.0\linewidth}{!}{
    \tablestyle{3.0pt}{1.00}
\begin{tabular}{lcccc}
\toprule
\multicolumn{1}{l|}{\textbf{Model Name}} & \multicolumn{1}{c|}{\textbf{Personalization}} & \multicolumn{1}{c|}{\textbf{Relevance}} & \multicolumn{1}{c|}{\textbf{Diversity}} & \textbf{Factuality} \\ \midrule
\rowcolor{COLOR_MEAN}
\multicolumn{5}{l}{\textit{Proprietary LLMs with Zero-shot Prompting}}                                                                                                                                     \\
\multicolumn{1}{l|}{GPT-5.2}             & \multicolumn{1}{c|}{6.6}                      & \multicolumn{1}{c|}{67.1\%}             & \multicolumn{1}{c|}{14.6\%}             & 55.3\%             \\
\multicolumn{1}{l|}{Gemini-3-Pro}        & \multicolumn{1}{c|}{5.3}                      & \multicolumn{1}{c|}{78.7\%}             & \multicolumn{1}{c|}{10.9\%}             & 49.0\%             \\ \hline
\rowcolor{COLOR_MEAN}
\multicolumn{5}{l}{\textit{Proprietary LLMs with Few-shot Prompting}}                                                                                                                                      \\
\multicolumn{1}{l|}{GPT-5.2}             & \multicolumn{1}{c|}{6.6}                      & \multicolumn{1}{c|}{69.3\%}             & \multicolumn{1}{c|}{14.3\%}             & 57.1\%             \\
\multicolumn{1}{l|}{Gemini-3-Pro}        & \multicolumn{1}{c|}{5.7}                      & \multicolumn{1}{c|}{79.1\%}             & \multicolumn{1}{c|}{11.3\%}             & 49.5\%             \\ \hline
\rowcolor{COLOR_MEAN}
\multicolumn{5}{l}{\textit{Open-source LLMs with Tuning-based Methods}}                                                                                                                                    \\
\multicolumn{1}{l|}{Qwen3-8B$_{\text{SFT}}$}       & \multicolumn{1}{c|}{6.8}                                           & \multicolumn{1}{c|}{78.1\%}                                  & 38.9\%                                  & 94.8\%             \\
\multicolumn{1}{l|}{Qwen3-8B$_{\text{RAG}}$}       & \multicolumn{1}{c|}{7.1}                                           & \multicolumn{1}{c|}{81.2\%}                                  & 42.2\%                                  & 97.4\%             \\ \midrule
\rowcolor{LightCyan}
\multicolumn{1}{l|}{\textbf{MuChator}}   & \multicolumn{1}{c|}{\textbf{8.4}}             & \multicolumn{1}{c|}{\textbf{89.1\%}}    & \multicolumn{1}{c|}{\textbf{51.1\%}}    & \textbf{99.3\%}    \\ \bottomrule
\end{tabular}
}
    \vspace{-0.6em}
\end{table}

\subsubsection{Offline Results.}
Table~\ref{main_result} reports the full offline results of our proposed method across four metrics. MuChator outperforms proprietary LLM and tuning-based baselines in personalization, relevance, diversity, and factuality, with the key findings summarized as follows:
\textbf{(1) Enhanced Personalization Quality}. MuChator achieves a personalization score of 8.4, whereas the fine-tuned Qwen3-8B$_{\text{RAG}}$ reaches only 7.1, indicating that MuChator's recommendations better align with user preferences.
\textbf{(2) Superior Music-Query Collaborative Relevance } Compared with the strongest closed-source baseline, Gemini-3-Pro, MuChator improves the relevance score from 78.7 to 89.1. These results indicate that MuChator can more effectively capture and align with users’ intent-driven queries.
\textbf{(3) Diverse Recommendation Outputs.} MuChator produces the most diverse recommendations, drawing from a broader range of music within the platform’s music catalog. Specifically, MuChator achieves a music coverage rate of 51.1\%, whereas all other models attain coverage rates of at most 42.2\%.
\textbf{(4) Faithful Music Recommendation.} MuChator achieves a substantially higher factuality score than other models. This indicates that MuChator can consistently recommend music that exists in the platform’s catalog, rather than hallucinating non-existent items.

\begin{table}[!ht]
  \centering
  \caption{Ablation studies on training stages with incremental inclusion of Context-aware Instruction Tuning (SFT), Music Knowledge Pre-training (Pretrain), and Preference Alignment with Hybrid RM (RL). }
    \vspace{-1.2em}
  \label{abla_training_stage}
    \resizebox{1.0\linewidth}{!}{
    \tablestyle{4.0pt}{1.0}
\begin{tabular}{ccc|c|c|c|c}
\toprule
\textbf{Pretrain} & \textbf{SFT} & \textbf{RL} & \textbf{Personalization} & \textbf{Relevance} & \textbf{Diversity} & \textbf{Factuality} \\ \midrule
\multicolumn{3}{c|}{\cellcolor{COLOR_MEAN}\textit{Vanilla Qwen3-8B}}                           & 6.1                      & 37.1\%             & 28.8\%             & 52.9\%              \\
                   & \checkmark         &             & 6.8 \textcolor{my_green}{($\uparrow$11.5\%)}                      & 78.1\% \textcolor{my_green}{($\uparrow$110.5\%)}             & 38.9\% \textcolor{my_green}{($\uparrow$35.1\%)}             & 94.8\% \textcolor{my_green}{($\uparrow$79.2\%)}              \\
\checkmark               & \checkmark         &             & 7.3 \textcolor{my_green}{($\uparrow$19.7\%)}                      & 87.8\% \textcolor{my_green}{($\uparrow$136.7\%)}             & \textbf{63.1\%} \textcolor{my_green}{($\uparrow$119.1\%)}             & 98.4\% \textcolor{my_green}{($\uparrow$86.0\%)}             \\
\checkmark               & \checkmark         & \checkmark        & \textbf{8.4} \textcolor{my_green}{($\uparrow$37.7\%)}                      & \textbf{89.1\%} \textcolor{my_green}{($\uparrow$140.2\%)}             & 51.1\% \textcolor{my_green}{($\uparrow$77.4\%)}             & \textbf{99.3\%} \textcolor{my_green}{($\uparrow$87.7\%)}              \\ \bottomrule
\end{tabular}
}
    \vspace{-1.1em}
\end{table}

\subsection{Ablation Studies}

\subsubsection{Ablation on Training Stage.}
To demonstrate the necessity of each training stage, we conduct ablation studies by progressively adding Context-aware Instruction Tuning (SFT), Music Knowledge Pre-training (Pre-train), and Preference Alignment with Hybrid RM (RL). As shown in Table~\ref{abla_training_stage}, adding pre-training to SFT substantially improves personalization, relevance, diversity, and factuality. Notably, the relevance score rises from 78.1 to 87.8, suggesting that pre-training strengthens the model’s ability to align recommendations with user intent. Moreover, RL improves personalization from 7.3 to 8.4 and relevance from 87.8 to 89.1, but reduces diversity, suggesting that RL stage prioritizes items aligned with user preferences and query intent, while limiting long-tail exploration.

\begin{table}[!t]
  \centering
  \caption{
  Ablation studies on pre-training stages with incremental inclusion of Stage 1, Stage 2, and Stage 3. Mixed pre-training denotes three-stage mixed training.
  }
    \vspace{-1.1em}
  \label{abla_pretrain_stage}
    \resizebox{1.0\linewidth}{!}{
    \tablestyle{4.0pt}{1.0}
\begin{tabular}{ccc|c|c|c|c}
\toprule
\textbf{\begin{tabular}[c]{@{}c@{}}Pre-train\\ Stage 1\end{tabular}} & 
\textbf{\begin{tabular}[c]{@{}c@{}}Pre-train\\ Stage 2\end{tabular}} & 
\textbf{\begin{tabular}[c]{@{}c@{}}Pre-train\\ Stage 3\end{tabular}} & 
\textbf{\begin{tabular}[c]{@{}c@{}}Music ($\uparrow$)\\ Knowledge \end{tabular}} & 
\textbf{\begin{tabular}[c]{@{}c@{}}Q2I ($\uparrow$)\\ Relevance \end{tabular}} & 
\textbf{\begin{tabular}[c]{@{}c@{}}U2I ($\downarrow$)\\ Personalization\end{tabular}} & 
\textbf{\begin{tabular}[c]{@{}c@{}}General ($\uparrow$)\\ Knowledge\end{tabular}} \\ 
\midrule

\multicolumn{3}{c|}{\cellcolor{COLOR_MEAN}\textit{Vanilla Qwen3-8B Inference}} 
& 61.1 
& 75.4 
& 10.51 
& \textbf{77.1} \\

\checkmark 
& 
& 
& 74.6 \textcolor{my_green}{($\uparrow$22.1\%)} 
& 84.6 \textcolor{my_green}{($\uparrow$12.2\%)} 
& 8.43 \textcolor{my_green}{($\downarrow$19.8\%)} 
& 76.4 \textcolor{my_red}{($\downarrow$0.9\%)} \\

\checkmark 
& \checkmark 
& 
& 80.4 \textcolor{my_green}{($\uparrow$31.6\%)} 
& 91.2 \textcolor{my_green}{($\uparrow$21.0\%)} 
& 8.12 \textcolor{my_green}{($\downarrow$22.7\%)} 
& 75.3 \textcolor{my_red}{($\downarrow$2.3\%)} \\

\checkmark 
& \checkmark 
& \checkmark 
& \textbf{82.4} \textcolor{my_green}{($\uparrow$34.9\%)} 
& \textbf{93.2} \textcolor{my_green}{($\uparrow$23.6\%)} 
& \textbf{3.64} \textcolor{my_green}{($\downarrow$65.4\%)} 
& 74.3 \textcolor{my_red}{($\downarrow$3.6\%)} \\

\midrule

\multicolumn{3}{c|}{Mixed Knowledge Pre-training} 
& 78.8 \textcolor{my_green}{($\uparrow$29.0\%)} 
& 86.1 \textcolor{my_green}{($\uparrow$14.2\%)} 
& 4.15 \textcolor{my_green}{($\uparrow$60.5\%)} 
& 69.2 \textcolor{my_red}{($\downarrow$10.2\%)} \\

\bottomrule
\end{tabular}
}
 \vspace{-0.5em}
\end{table}

\begin{table}[!t]
  \centering
  \caption{Ablation on User Context by  with incremental inclusion of item, user profile, state, and feedback.}
     \vspace{-1.1em}
  \label{abla_u2i}
    \resizebox{1.0\linewidth}{!}{
    \tablestyle{2.0pt}{0.7}
\begin{tabular}{cccc|c}
\toprule
\textbf{\begin{tabular}[c]{@{}c@{}}User Profile\end{tabular}} & 
\textbf{\begin{tabular}[c]{@{}c@{}}Item\end{tabular}} & 
\textbf{\begin{tabular}[c]{@{}c@{}}State\end{tabular}} & 
\textbf{\begin{tabular}[c]{@{}c@{}}Feedback\end{tabular}} & 
\textbf{\begin{tabular}[c]{@{}c@{}} U2I Personalization  ($\downarrow$)\end{tabular}}  \\ 
\midrule

       & \checkmark        &     &   &  4.81 \\

\checkmark      &    \checkmark    &  &   &  4.46 \textcolor{my_green}{($\downarrow$7.3\%)} \\

\checkmark      &    \checkmark    &  \checkmark  &  &  4.13 \textcolor{my_green}{($\downarrow$14.1\%)} \\

\checkmark      &    \checkmark    &  \checkmark  &  \checkmark  &  \textbf{3.64} \textcolor{my_green}{($\downarrow$24.3\%)} \\

\bottomrule
\end{tabular}
}
 \vspace{-0.5em}
\end{table}

\begin{table}[!t]
  \centering
  \caption{Ablation studies on hybrid rewards by individually removing the personalization (Pers.), relevance (Rel.), and rule-based (Rule.) rewards.}
    \vspace{-1.1em}
  \label{abla_reward}
    \resizebox{1.0\linewidth}{!}{
    \tablestyle{4.0pt}{1.0}
\begin{tabular}{ccc|c|c|c|c}
\toprule
\textbf{Rel.} & \textbf{Pers.} & \textbf{Rule.} & \textbf{Personalization} & \textbf{Relevance} & \textbf{Diversity} & \textbf{Factuality} \\ \midrule
\checkmark               & \checkmark                     & \checkmark                & \underline{8.4}                      & \underline{89.1}\%             & \textbf{51.1\%}    & \textbf{99.3\%}     \\
\checkmark               &\crossmark                          & \checkmark                & 6.8                      & \textbf{90.5\%}    & \underline{45.7\%}             & \underline{99.1\%}              \\
\crossmark                   & \checkmark                     & \checkmark                & \textbf{10.5}            & 43.0\%             & 27.4\%             & 98.6\%              \\
\checkmark               & \checkmark                     & \crossmark                    & 8.1                      & 88.5\%             & 48.4\%             & 97.2\%              \\ \bottomrule
\end{tabular}
}
 \vspace{-1.0em}
\end{table}

\subsubsection{Ablation on Pre-training Stage.}
We further conduct ablations to examine the effect of each music knowledge pre-training stage: (1) vanilla Qwen3-8B inference; (2) Stage 1 with objective music knowledge; (3) Stage 2 with subjective music knowledge; (4) Stage 3 with personalized music preferences; and (5) mixed knowledge pre-training. We evaluate all models on four benchmarks: Music Knowledge, Q2I Relevance, U2I Personalization, and General Knowledge. We report multiple-choice accuracy for Music Knowledge, Q2I Relevance, and General Knowledge, while for U2I Personalization, we report perplexity of the generated songs. The results are reported in Table~\ref{abla_pretrain_stage}. Introducing knowledge at different pre-training stages consistently improves performance across the Music Knowledge, Q2I Relevance, and U2I Personalization benchmarks. Notably, Stage 3 substantially improves U2I Personalization, reducing PPL from 8.12 to 3.64, highlighting the importance of personalized preference modeling for aligning recommendations with user behavior. We also compare curriculum pre-training with mixed pre-training and find that the curriculum strategy performs consistently better. This can be attributed to the fact that curriculum learning enables an easy-to-hard learning process, in which the model first acquires music knowledge before learning user listening sequences, leading to more effective representation learning.

\subsubsection{Ablation on User Context Sequence.}
To analyze the effect of each component in the time-series user context, we progressively incorporate the music item prediction task with user profiles, user states, and user feedback. 
As shown in Table~\ref{abla_u2i}, incorporating different user context components consistently improves U2I Personalization, with the PPL decreasing from 4.81 to 3.64. This result indicates that richer user context provides more informative signals for modeling user preferences.

\subsubsection{Ablation on Hybrid Rewards.}
We evaluated the effectiveness of different rewards in the RL stage, and the results are shown in Table~\ref{abla_reward}. For both relevance and personalization rewards, removing any individual reward component leads to a degradation in the corresponding metric. For example, removing the relevance reward reduces the relevance score from 89.1 to 43.0. In addition, incorporating rule-based rewards to penalize non-existent and repeated music effectively improves diversity and factuality.

\subsection{Experimental Analysis}

\begin{figure}[!t]
  \centering
    \includegraphics[width=0.98\linewidth]{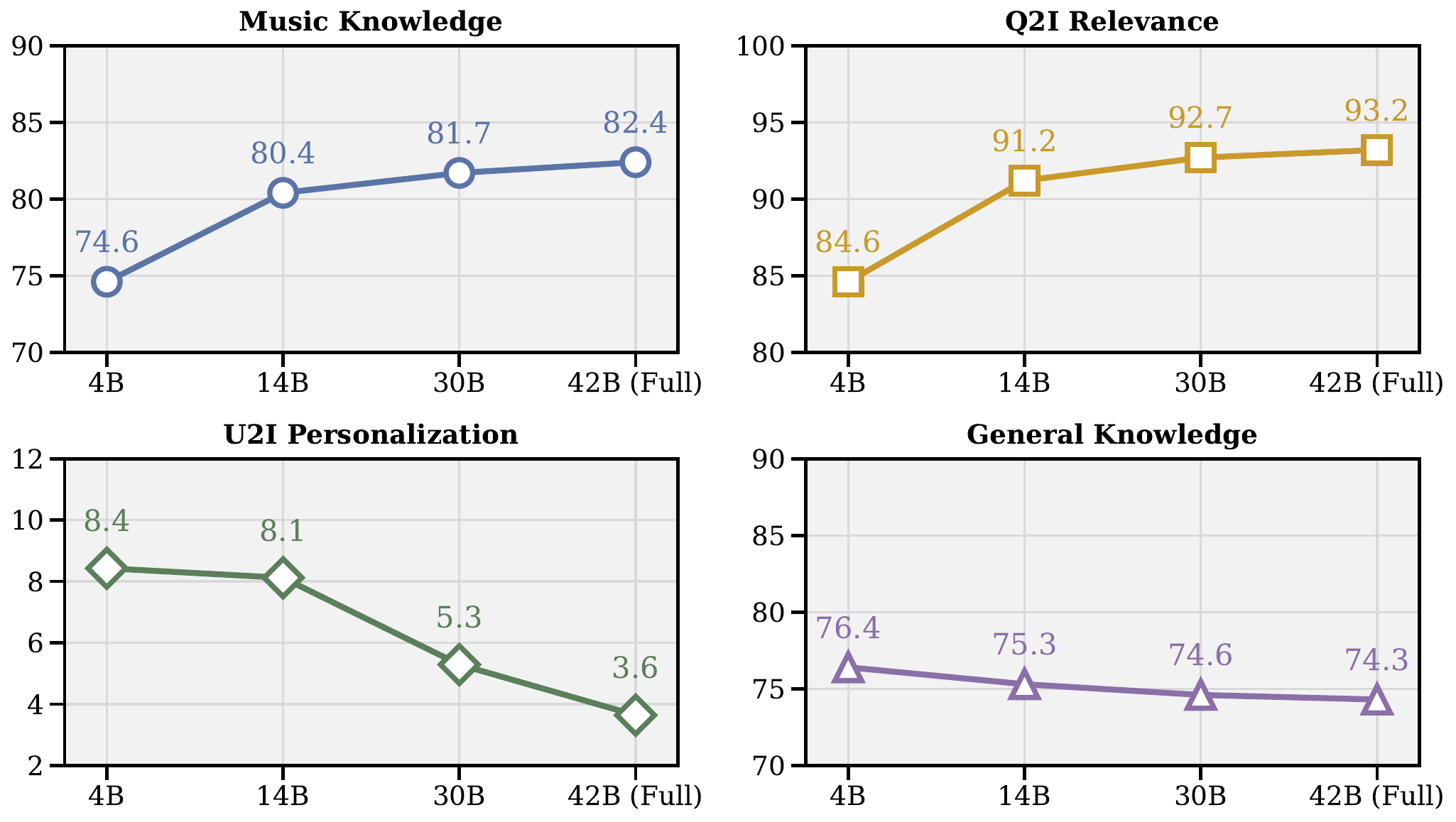}
    \vspace{-1.1em}
    \caption{ Impact of data size on pre-training performance across four tasks: Music Knowledge, Q2I Relevance, U2I Personalization, and General Knowledge.
    }
    \vspace{-1.0em}
    \label{fig:model_size}
\end{figure}

\subsubsection{Analysis on Data Scaling.}
We evaluate the impact of token scale on pre-training performance on four tasks: music knowledge, Q2I relevance, U2I personalization, and general knowledge. Scaling curves are presented in Figure~\ref{fig:model_size}. Increasing the token budget consistently improves performance on the three music-related tasks, with only slight degradation on general knowledge, indicating that domain-specific pre-training enhances music-domain capabilities without substantially compromising generality.

\subsubsection{Analysis on Complex Queries.}
To evaluate MuChator’s ability to handle complex user intents, we categorize real user queries in an online environment and track the proportion of complex versus simple queries. The results show that after deploying MuChator, complex queries increases by 12.3\%, suggesting that incorporating MuChator enables the system to better satisfy users’ complex query needs, thereby encouraging users to issue more such queries.

\section{Conclusion}
In this paper, we introduce MuChator, an interactive MusicLLM-based framework that enables users to actively express situational music intents in natural language.
By introducing a three-stage Music Knowledge Pre-training, MuChator incrementally injects LLMs with objective music knowledge, subjective music knowledge, and personalized music preferences. 
To model intent relevance and personalized preferences, we introduce a Preference Alignment Post-training that constructs User–Query-to–Music triplets for instruction tuning, and optimizes hybrid rewards via GRPO-based reinforcement learning.
Online and offline evaluations demonstrate the effectiveness of MuChator in improving user experience and platform revenue, with a 46.49\% improvement on user active days.
The assistant has been deployed on Douyin Music App.


\bibliographystyle{ACM-Reference-Format}
\bibliography{sample-base}


\end{document}